\documentclass[journal=jacsat,manuscript=article]{achemso}

\usepackage{chemformula} 
\usepackage[T1]{fontenc} 

\usepackage[caption=false]{subfig} 

\author{Kostas Kanellopulos}
\author{Robert G. West}
\author{Silvan Schmid}
\email{silvan.schmid@tuwien.ac.at}
\affiliation[TU Wien]
{Institute of Sensor and Actuator Systems, TU Wien, Gusshausstrasse 27-29, 1040 Vienna, Austria.}

\title{Nanomechanical Photothermal Near Infrared Spectromicroscopy of Individual Nanorods}
  
\abbreviations{NEMS, NIR, FEM, SPR}

\begin{document}

\begin{abstract}
Understanding light-matter interaction at the nanoscale requires probing the optical properties of matter at the individual nano-absorber level. To this end, we have developed a nanomechanical photothermal sensing platform that can be used as a full spectromicroscopy tool for single molecule and single particle analysis. As a demonstration, the absorption cross-section of individual gold nanorods is resolved from the spectroscopic and polarization standpoint. By exploiting the capabilities of nanomechanical photothermal spectromicroscopy, the longitudinal localized surface plasmon resonance (LSPR) in the NIR range is unravelled and quantitatively characterized. The polarization features of the transversal surface plasmon resonance (TSPR) in the VIS range are also analyzed. The measurements are compared with the finite element method (FEM), elucidating the role played by electron-surface and bulk scattering in these plasmonic nanostructures, as well as the interaction between the nano-absorber and the nanoresonator, ultimately resulting in absorption strength modulation.
Finally, a comprehensive comparison is conducted, evaluating the signal-to-noise ratio of nanomechanical photothermal spectromicroscopy against other cutting-edge single molecule and particle spectroscopy techniques. This analysis highlights the remarkable potential of nanomechanical photothermal spectromicroscopy due to its exceptional sensitivity.
\end{abstract}

\section{Introduction}

The advent and development of optical single-molecule and single-particle measurement techniques have had a tremendous impact on the scientific research for the past 30 years \cite{Moerner2020}. The detection of single objects at the nanoscale level affords us a unique perspective on the interactions occurring between these tiny entities and their local environment, revealing their heterogeneity without relying on ensemble average information.
Due to its high signal-to-noise ratio (SNR), optical fluorescence-based detection approaches have rapidly evolved, being nowadays routinely employed in a huge variety of scientific fields, from biology to condensed matter, to the design and engineering of novel materials. However, the fluorescent label can photobleach or quench as well as alter the system under study. For this reason, the scientific community has pushed toward the development of label-free single-molecule detection schemes \cite{Arroyo2016}, such as iSCAT \cite{Celebrano2011, Priest2021}, ground-state depletion microscopy \cite{Chong2010} and photothermal microscopy \cite{Adhikari2020, Chien2018}, among others.
All these approaches rely on the absorption rather than scattering of the nano-object upon illumination of a probing light. The rationale behind this choice lies in the fact that, while the optical scattering cross-section scales quadratically with the target volume ($\sigma_{scatt}$ $\propto$ $V^2$), the optical absorption cross-section scales linearly with it ($\sigma_{abs}$ $\propto$ $V$) \cite{Bohren1983, Baffou2013}. In other words, the smaller the target size is, the more effectively its absorption properties can be interrogated. More specifically, this central aspect shows also the advantage of nanomechanical absorption spectroscopy over other fully-optical single-molecule spectroscopic methods, such as surface-enhanced Raman scattering (SERS) \cite{Langer2020} or tip-enhanced Raman spectroscopy (TERS) \cite{Pienpinijtham2022}. The former approach uses the strong near-field enhancement at the nanoscale on the surface of plasmonic nanoparticles or nanostructures (so-called hotspots) to increase the SNR of the Stokes-shift Raman scattering. However, the plasmonic nanostructure fabrication and control of the placement of the particles/molecules in the sites of interest increase the overall complexity of the measurement procedure. The latter approach uses a nanoscopic probe to scan an area where the molecules of interest are fixed, which requires a very good control in the tip fabrication. In contrast, nanomechanical absorption spectroscopy overcomes all this complexity as it measures directly the non-radiative energy losses of the illuminated molecule, not limited to any specific sample preparation. In other words, the molecule itself becomes part of the detector, due to the interplay between its absorption properties and the light excitation used. The reduced analyte-detector distance results also in a reduction of noise and unwanted external interference. Based on this consideration, it has been possible to detect and image single molecules by nanomechanical photothermal sensing.\cite{Chien2018} This work has been made possible by the previous research, which showed the ability of nanomechanical resonators to detect and quantify the absorption of single plasmonic \cite{Schmid2014,Ramos2018,chien2021analysis} and polymer \cite{Larsen2013} nanoparticles via photothermal heating. It is worth noting that this photothermal spectromicroscopy approach is not based on the thermo-optical effect as in photothermal contrast microscopy \cite{Gaiduk2010, Chang2012, Ding2016, Hou2017, Adhikari2020}, where the absorber is detected due to the temperature dependence of the surrounding embedding medium refractive index (glycerol, thermotropic liquid crystal, near-critical $Xe$ or $CO_2$) \cite{Adhikari2020, Wang2023} via modulation of the scattering of a second probing laser. In nanomechanical photothermal spectromicroscopy, this thermal effect consists instead in a stress reduction in the nanomechanical resonator, detuning its resonance frequency upon illumination of the nano-absorber.

Here, this work pushes further the boundaries of single-molecule nanomechanical photothermal sensing toward a full NIR spectro- and polarization-microscopy technique. With a silicon nitride nano-optomechanical drum resonator as a sensitive thermometer, individual gold nanorods are localized and their spectra and polarization features fully characterized, additionally shedding light on their interaction with the nanoresonator itself. Among the huge variety of nanoparticle shapes and materials, gold nanorods occupy a relevant position in gas- and liquid-phase chemical detection, so as sensing platform for biomolecules \cite{Cao2014, Nooteboom2022}, or as photothermal heating sources \cite{Shukla2020}. In this work, the plasmonic properties of such nano-objects are analyzed and their corresponding plasmonic damping mechanisms unravelled, showing also a good agreement with finite element method (FEM) simulation results. The performance of our approach is then compared with the other state-of-the-art single molecule and particle techniques in terms of normalized SNR, showing the capabilities offered by nanomechanical photothermal spectromicroscopy with its superior SNR. 

\section{Experimental methods}

\begin{figure*}
    \begin{center}
    \centering
    \captionsetup{justification=justified}
    \includegraphics[width = 1\textwidth]{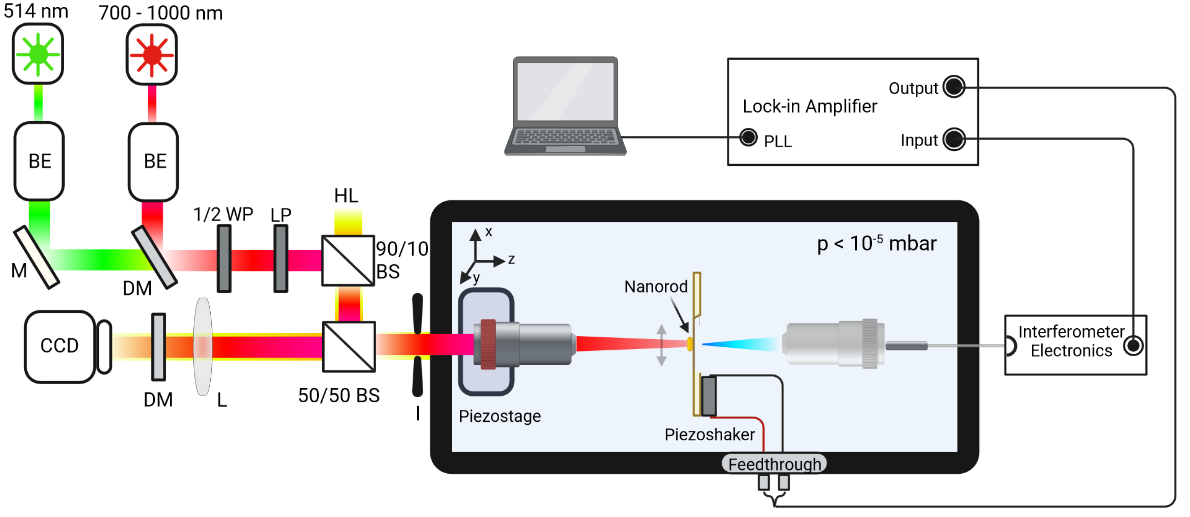}
    \caption{Schematics of the operating set-up. The drum resonator is actuated in vacuum ($p$ $<$ $10^{-5}$ $mbar$) by a piezoshaker. The displacement is read out by a Fabry-Perot interferometer (blue laser). The interference signal is processed and sent to the lock-in amplifier which records the frequency. The scanning lasers (red and green) are used to generate the photothermal signal by plasmonic excitation of the nanorod. BE: beam expander. BS: beam splitter. CCD: charge-coupled device camera. DM: dichroic mirror. I: iris. L: lens. LP: linear polarizer. HL: halogen lamp. M: mirror. WP: waveplate.}
    \label{fig:setup}
    \end{center}
\end{figure*}

The 50 nm thick nano-optomechanical drum resonator is operated  at room temperature under high vacuum conditions ($p$ < $10^{-5}$ mbar) in order to reduce air damping and eliminate heat dissipation by convection \cite{Chien2018}. It is actuated by a piezo-electric element placed underneath the sample holder. The nanoresonator's displacement is measured with a Fabry-Perot interferometer (Attocube IDS3010) \cite{Thurner2013, Thurner2015} (see Fig.~\ref{fig:setup}). The mechanical resonance frequency is recorded with a phase-locked loop (PLL) tracking scheme (HF2LI, Zurich Instrument).

The operating set-up is equipped with a green laser at 513 nm (Toptica TopMode) and a Ti:Sapphire laser (M Square SolsTiS) with a tunable wavelength in the range of 700-1000 nm, used as probe beams to photothermally excite both the longitudinal (LSPR) and transverse plasmonic resonances (TSPR) of each nanorod. In fact, every time one of the two lasers is scanned across the central area of the drum and hits the sample, the corresponding light absorption results in a local heating, reducing the stress of the nanoresonator and ultimately resulting in a detectable resonance frequency detuning \cite{Chien2018, Schmid2014, Larsen2013}.

For the scanning laser probes, long working distance 50x objectives are used (N.A. = 0.42, M Plan Apo NIR, Mitutoyo in the NIR range, N.A. = 0.55, M Plan Apo, Mitutoyo in the VIS range). The lasers' polarization angle is controlled by means of a linear polarizer in the optical beam path. Raster scanning is made possible by a closed-loop piezoelectric nanopositioning stage (PiMars, Physikinstrumente). The analyte sampling is performed by spin-casting onto the resonator a drop of diluted solution containing the nanorods, after being filtered with 200 nm pore size PTFE membrane syringe filters (Acrodisc, Sigma-Aldrich) to avoid particle aggregation.
As already shown \cite{Chien2018}, NEMS-based photothermal sensing enables the measurement of pure optical absorption of the sample, resulting in the full characterization of its absorption cross-section
\begin{equation} \label{eq1}
    \sigma_{abs}(\lambda) = \frac{P_{abs}}{I_0},
\end{equation}
with $I_0$ being the peak irradiance of a gaussian laser beam and $P_{abs}$ the absorbed power by the sample. The former is a function of the input laser power $P_0$
\begin{equation} \label{eq2}
    I_{0} = \frac{2P_{0}}{\pi r^2},
\end{equation}
with $r$ being the beam radius, which is always characterized by knife-edge method before each measurement \cite{Khosrofian1983, Marshall2010}.
The latter can be calculated from the measured resonance frequency shifts, assuming a full thermalization, as
\begin{equation} \label{eq3}
    P_{abs} = \frac{\Delta f}{f_0 \ R_P},
\end{equation}
$f_0$ being the resonance frequency, $\Delta f$ the frequency shift, and $R_P$ the relative power responsivity ($W^{-1}$).

\section{Results and discussion}
\begin{figure}
  \begin{center}
  \centering
  \subfloat[]{\includegraphics[width = 0.50\textwidth]{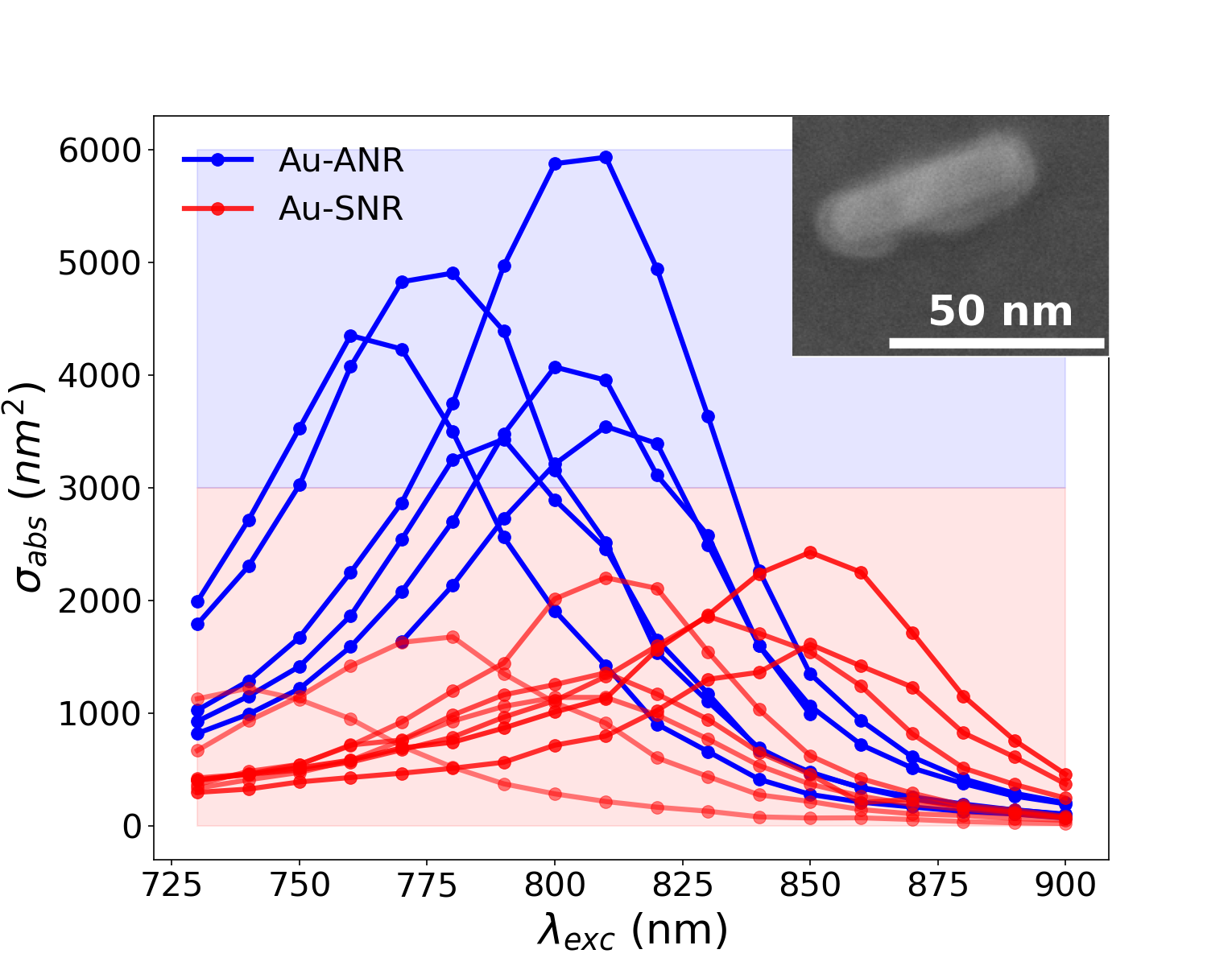} \label{fig:snr spectra}}
  \hfill
  \subfloat[]{\includegraphics[width = 0.50\textwidth]{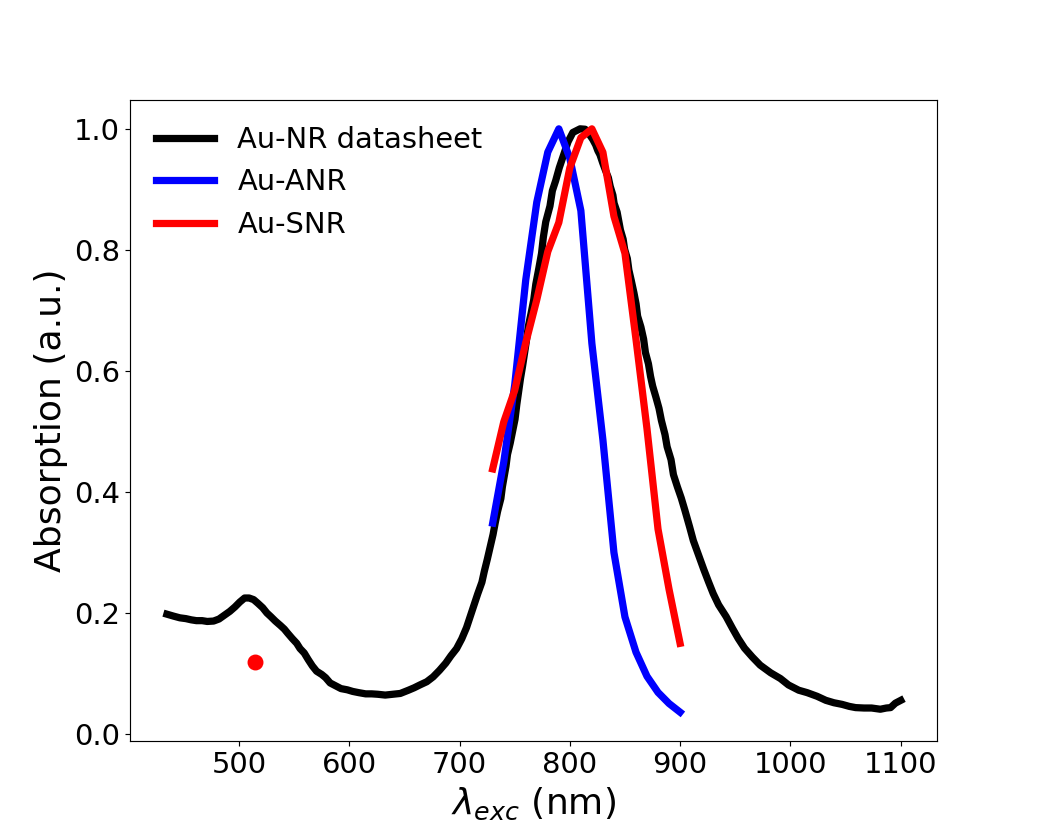} \label{fig:signal sum}}
  \caption{(a) Measured absorption cross-sections spectra of single nanorods (Au-SNR, red curves) and small nanorods aggregations (Au-ANR, blue curves), showing the heterogeneity characterizing these samples, mainly caused by the particle size dispersion. Inset: SEM micrograph of a single silica-coated gold nanorod landing on the drum resonator. (b) Red curve and dot: renormalized sum of the measured absorption cross-section spectra of single nanorods (fig.~\ref{fig:snr spectra}); blue curve: renormalized sum of the absorption cross-sections of the nanorods aggregation; black curve: ensemble average absorption spectrum given by the datasheet.} \label{fig:spectra}
  \end{center}
\end{figure}

The nanorods analyzed here have lengths in the range of ca. 38-48 nm, radial diameters in the range of 9.5-11.5 nm and silica coating with a thickness in the range of 18-22 nm (Sigma-Aldrich silica coated gold nanorods) (Fig.~\ref{fig:snr spectra}, inset - the majority of the nanorods have been SEM imaged by deposition of 10 nm thick gold layer on top of them to reduce any possible charging effect). Their optical properties in the visible and near-infrared range are characterized by surface plasmon resonances (SPR), i.e. the electromagnetic coupling between an impinging light and the collective motion of the conduction band electrons \cite{Bohren1983, Maier2007}. The main interest in gold nanorods lies in their large SPR amplitudes and broad spectral tunability \cite{Muskens2008, Myroshnychenko2008, Chang2010, Chen2011, Davletshin2012, Crut2014, Zilli2019, Heylman2016, Rangacharya2020, Brioude2005}. In the specific case where a coating is present, depending on its thickness, SPR features will be more or less affected by the environment. Indeed, the plasmonic response is sensitive to the surrounding on the spatial range of the order of the nanorod diameter, the region where the field enhancement takes place \cite{Davletshin2012, Liz2008, Lioi2022, Tian2012} (Supporting Information Fig.~S1). In the present study, a silica coating of roughly 20 nm is thin enough for the SPR to remain sensitive to both the coating and the surrounding medium, but within a reduced magnitude \cite{Lioi2022}.

\begin{figure}
  \begin{center}
  \centering
  \subfloat[]{\includegraphics[width = 0.50\textwidth]{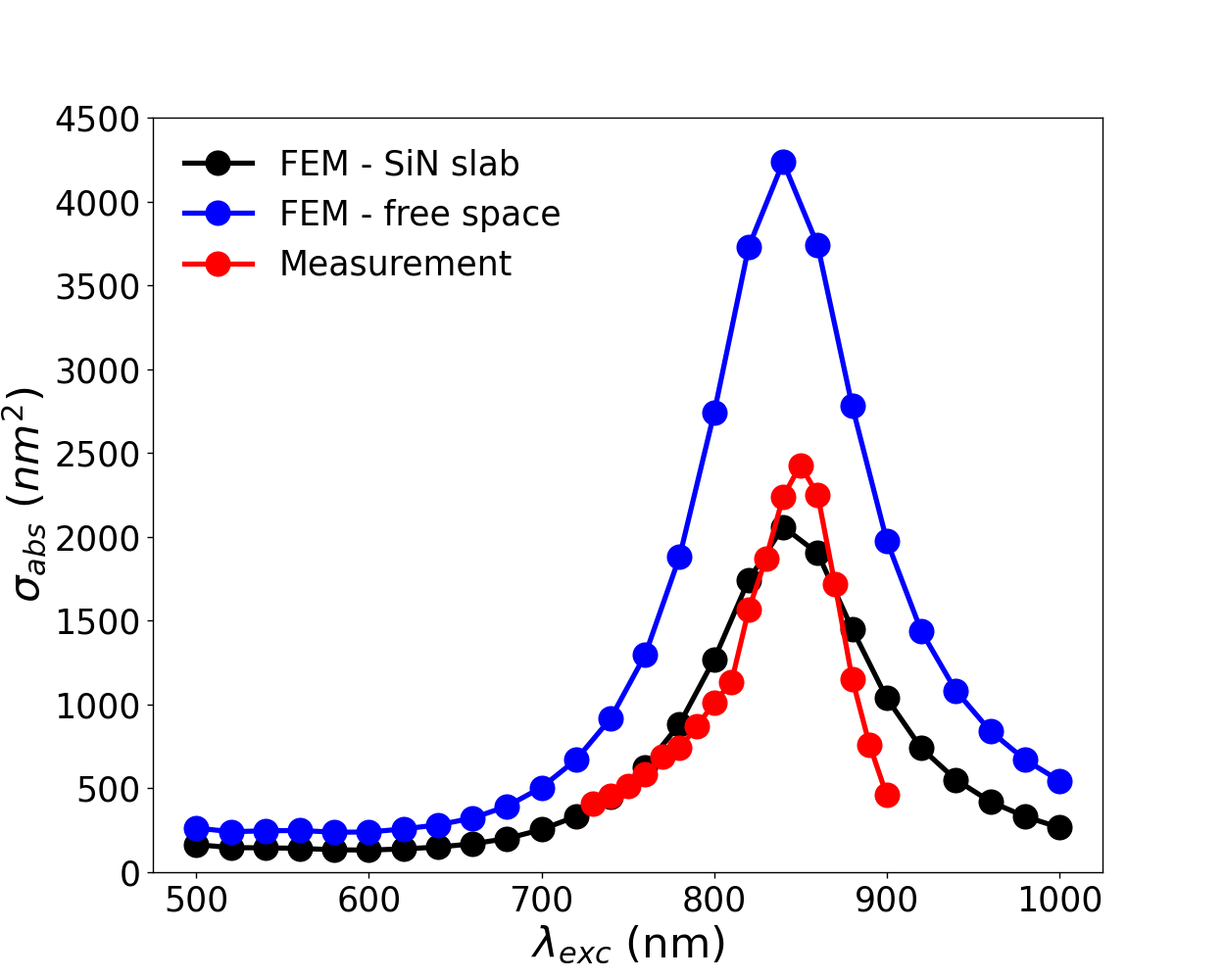} \label{fig:spectrum snr}}
  \hfill
  \subfloat[]{\includegraphics[width = 0.50\textwidth]{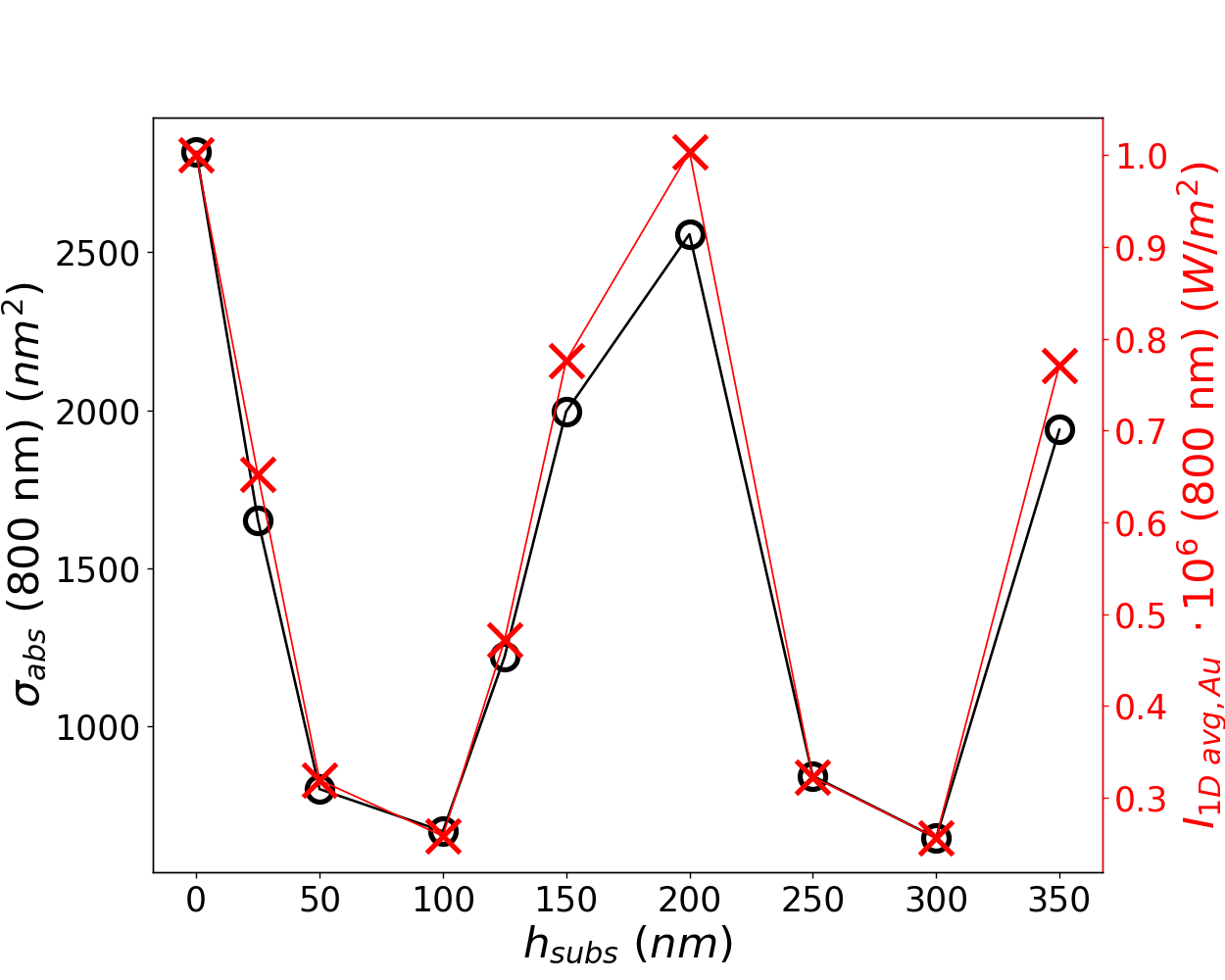} \label{fig:acs I substrate}}
  \caption{(a) Measured absorption cross-section of a individual nanorod (red dots), compared to FEM simulated absorption spectrum in the presence of the silicon nitride substrate (black dot) and in free space (blue dots), obtained for nanorod dimensions of L = 48 nm, r = 6 nm, with the silica coating thickness of 20 nm. (b) FEM absorption cross-section at 800 nm wavelength for the same nanorod (black empty dot and solid line), and 1D averaged FEM electromagnetic intensity in the vicinity of the gold core in the presence of the substrate only (red crosses and solid line), for different silicon nitride slab thicknesses. } \label{fig:single nr}
  \end{center}
\end{figure}

Fig.~\ref{fig:snr spectra} shows the measured absorption spectra of different, individual silica-coated gold nanorods (Au-SNR) indicated by the red curves, together with some aggregations of few units (Au-ANR) indicated by the blue curves. Differentiating these two photothermal responses is aided by FEM electromagnetic simulations (for an overview on how aggregations can be differentiate from individual nanorods, see Supporting Information). For each spectrum, the polarization of the probe beam (here Ti:Sapphire laser) was set to maximize the absorption in the wavelength range of 700-900 nm. In fact, the nanorods analyzed here present maximum absorption in the range of ca. 790-830 nm (see Fig.\ref{fig:signal sum}, black solid curve), due to LSPR excitation whenever the laser polarization is parallel to their long axis. With the nanomechanical photothermal technique, nanoparticle heterogeneity can be investigated, revealing more than the information obtained in ensemble average measurements. Here, the heterogeneity is mainly due to the size dispersion of the particles, as stated by the vendor \cite{DatasheetNR}, and it relates both to the LSPR spectral position $\lambda_{LSPR}$ and the absorption cross-section amplitude $\sigma_{abs}(\lambda_{LSPR})$. Moreover, the latter has a strong dependence on the substrate, specifically on its thickness and optical properties (see Fig.~\ref{fig:single nr}; further details in the Supporting Information). 

The red solid line in Fig.~\ref{fig:signal sum} indicates the corresponding renormalized sum of the responses from individual nanorods (Au-SNR in Fig.~\ref{fig:snr spectra}), showing a very good match with the reference spectrum given in the datasheet (black solid line), recovering a typical ensemble average absorption spectrum \cite{Muskens2008, Chang2010}.  The ensemble Au-SNR wavelength $\lambda_{LSPR}$ is measured to be 809~nm, which is close to the nominal value of 808~nm \cite{DatasheetNR}. Fig.~\ref{fig:signal sum} also shows the renormalized spectrum considering only responses from nanorod aggregations (Au-ANR, blue solid line). For this spectrum, an Au-ANR wavelength of 786~nm is extracted, corresponding to a blue-shift of the 2.8$\%$ from the Au-SNR wavelength. As shown by Jain et al. \cite{Jain2006}, such a blue-shift occurs in nanorod aggregations of two or more units assembled in a side-by-side orientation, for a polarization parallel to their long axis. As the authors reported, the strength of this shift depends on the interdistance between the nanorods involved, on their aspect ratios, the relative orientational angle and on the number of units considered. For the spectral distributions seen in Fig.~\ref{fig:spectra} (blue curves), we expect these signals to originate from side-by-side assembled nanorod aggregations. 

The individual nanorods have also been measured with a wavelength of 513~nm, to excite the TSPR (red dot in Fig.~\ref{fig:signal sum}). As expected, the absorption at this wavelength is roughly one order of magnitude smaller than LSPR, due to the overlap between the transverse localized plasmonic resonance and the electronic interband transitions, which start to arise at 2.4~eV ca. in gold \cite{Johnson1972, Kolwas2020, Sonnichsen2002}, ultimately resulting in enhanced plasmonic damping.


Fig.~\ref{fig:spectrum snr} focuses the attention on the nanomechanical photothermal spectrum of a individual nanorod (red dots). This specific sample has maximum absorption at $\lambda \approx 840$ nm with a cross-section of $\sigma_{abs} \approx 2.5 \cdot 10^{-15}$ $m^2$, close to what it is reported in literature \cite{Crut2014, Chen2011, Davletshin2012}. The measurements are compared with FEM simulations (black and blue dots, Fig.~\ref{fig:spectrum snr}), showing a good agreement with data (black dots). Indeed, FEM approaches offer the possibility to evaluate the absorption and scattering cross-sections of an arbitrarily shaped particle \cite{Parsons2010, Loke2014, Nima2017, Grand2020, Leyva2021, Knight2009, Comsol2022}. For this individual nanorod, a linewidth of ca. 132 meV is measured in terms of full-width-at-half-maximum (FWHM), two time higher than expected for electron-bulk scattering alone (72 meV, see Supporting Information). Intrinsic size effects for the gold core have to be taken into account, such as electron-surface scattering and radiative damping (Supporting Information Eq.~S3). The former affects the gold nanorods' LSPR linewidth stronger than the latter due to their reduced volume, in contrast to what has been observed in spherical gold nanoparticles \cite{Sonnichsen2002}. Similar FWHM values are found for the other nanorods, corroborating the evidence that electron-surface scattering is a major source of damping in these plasmonic-assisted nano-absorbers (Supporting Information Fig.~S3b).

Fig.~\ref{fig:spectrum snr} also clearly shows how the presence of the silicon nitride resonator affects ultimately the absorption cross-section of the nanorod under study. The FEM analysis in the presence of the substrate (black dots) reproduces well the measured absorption spectrum, where the FEM analysis conducted in free space in the absence of the slab does not (blue dots). More precisely, at the LSPR wavelength (840 nm), the absorption cross-section results are roughly half of the free space case ($\approx 2 \cdot 10^{-15}$ $m^2$ in the presence of the slab and $\approx 4.2 \cdot 10^{-15}$ $m^2$ in air). In contrast, the LSPR wavelength and FWHM are weakly affected by the presence of the substrate ($\leq$ 1$\%$ difference for both quantities). In general, a dielectric substrate underneath a metal nanoparticle screens the electromagnetic restoring force acting on the plasmon oscillations. This screening can be modelled qualitatively as a nanoparticle image with a reduced number of charges, whose electromagnetic strength is determined by the nanoparticle-substrate interdistance and the slab dielectric permittivity \cite{Knight2009}. The interdistance of 20 nm (silica coating thickness) and the relatively small refractive index of low-stress silicon nitride (whose spectral distribution has been taken from Ref.~\citenum{Philipp_1973}) give reason for this weak effect \cite{Lioi2022, Knight2009}. To better understand the role played by the silicon nitride slab, FEM simulations have been performed at a single wavelength (800 nm) for different thicknesses $h_{subs}$.

In Fig.~\ref{fig:acs I substrate} the absorption cross-section follows a period pattern for an increasing substrate thickness. This modulation follows perfectly the variation in intensity at the interface air-silicon nitride in the vicinity of the nanorod. In fact, the electromagnetic losses $Q_h$ due to absorption are directly proportional to the intensity of the electric field, $Q_h \ \propto \ |\bf{E}(\bf{r})|^2$ (Supporting Information Eq.~S1). This intensity modulation is due to the interference occurring between the input electric field and the reflected light from the slab, whose magnitude depends on the thickness and refractive index. The intensity value calculated here is an average over its spatial distribution in the proximity of the nanorod, but without it, in the presence of the substrate only (Supporting Information Fig.~S4). For the thickness used in this study (50 nm) along with the considered wavelengths (730-900 nm) and the refractive index of silicon nitride \cite{Philipp_1973}, no interference inside the slab is present as it is the case in Ref.~\citenum{Kosaka2014}. There, 1 $\mu m$ thick silicon cantilevers worked as optical cavity for specific wavelengths in the VIS range, modulating the scattering of deposited 100 nm gold nanoparticles. In this study, however, the most relevant interference occurs at the interface between the free space and the substrate. Therefore, by controlling the substrate refractive index and thickness, it is possible to tailor the absorption spectrum of individual nano-absorbers.

\begin{figure}
    \begin{center}
    \centering
    \includegraphics[width = 0.50\textwidth]{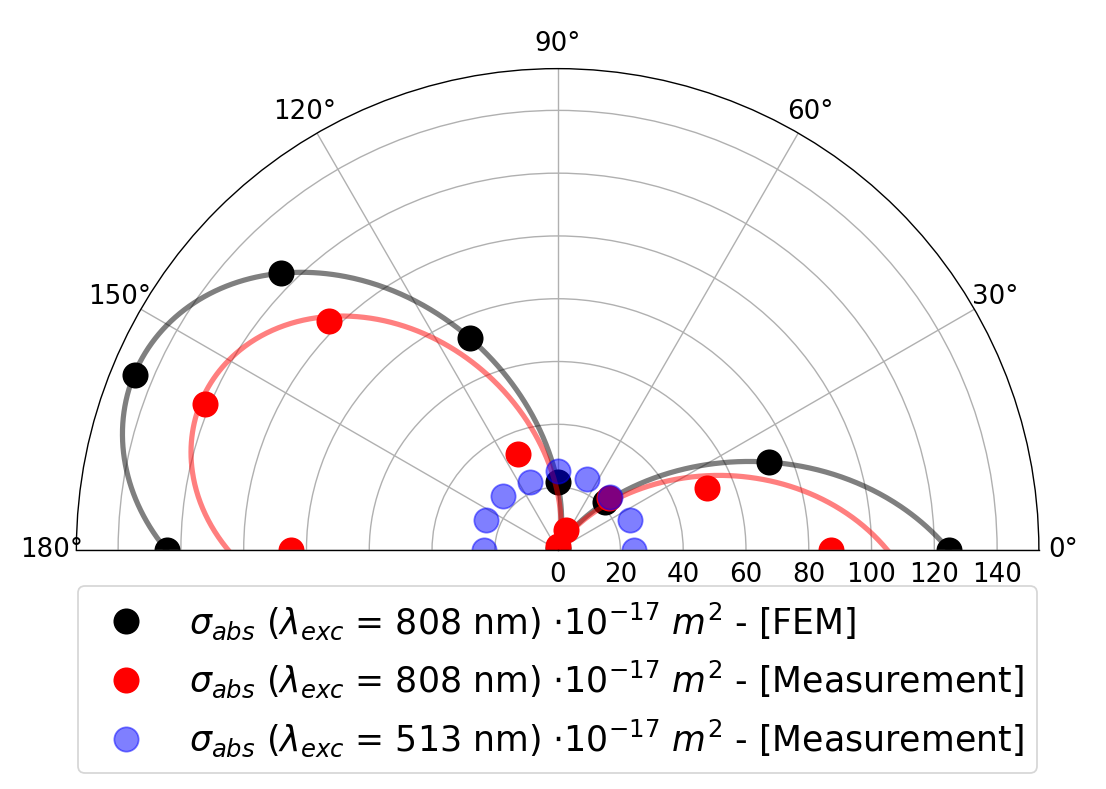}
    \caption{Polar plot of the absorption cross-section of an individual nanorod measured at $\lambda_{exc} = 808$ nm as a function of the polarization angle $\theta_{pol}$ (red dots). The ratio between the absorption cross-section for a polarization parallel to the nanorod long-axis ($\theta_{pol} \approx 157.5$°) and perpendicular to it ($\theta_{pol} \approx 90$°) is roughly $\sigma_{abs, \parallel}$(808 nm)/$\sigma_{abs, \perp}$(808 nm) $\approx 100$. FEM simulations show good agreement with the measurement (black dots). Both the red and black solid curves represent the $cos^2(\theta)$ pattern. Blue dots represent nanomechanical photothermal measurements at $\lambda_{exc}$ = 513 nm.}
    \label{fig:polar plot snr}
    \end{center}
\end{figure}

\begin{figure}
    \begin{center}
    \centering
    \includegraphics[width = 0.35\textwidth]{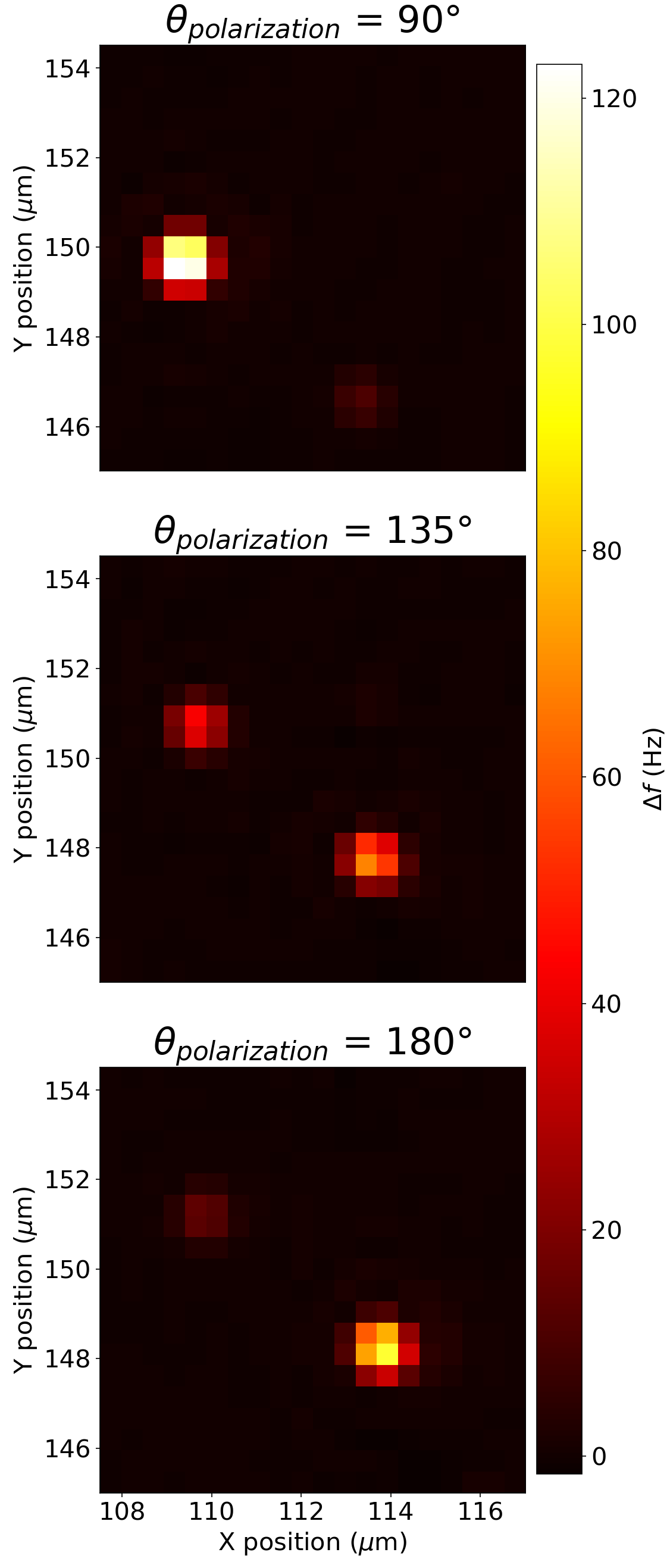}
    \caption{2D maps of the same region at $\lambda_{exc} = 808$ nm for three different polarization angles $\theta_{pol}$: 90°, 135°, 180°. The two responses are from two individual nanorods. For the perpendicular polarizations, 90° and 180°, the absorbers behave in an opposite way, meaning that they are almost perpendicular one to each other, while absorbing almost the same amount of light for the central map ($\theta_{pol} = 135$°).}
    \label{fig:2D scans}
    \end{center}
\end{figure}

For these non-spherical nanoparticles, absorption is strongly polarization dependent, as clearly seen in Fig.~\ref{fig:polar plot snr}\&\ref{fig:2D scans}. Fig.~\ref{fig:polar plot snr} shows how the absorption cross-section varies with the laser polarization angle for a individual nanorod, with the red dots representing the nanomechanical photothermal measurements close to its LSPR ($\lambda_{exc} = 808$ nm). Each point is acquired by changing the polarization of the scanning laser probe with steps of 22.5° by means of a half-waveplate (HWP) and a linear polarizer, while maintaining the same laser input power. The ratio between the absorption cross-section for a polarization parallel to the nanorod long-axis ($\theta_{pol} \approx 157.5$°) and perpendicular to it ($\theta_{pol} \approx 90$°) is roughly:
\begin{equation} \label{eq4}
    \frac{\sigma_{abs,\parallel}(808 \ nm)}{\sigma_{abs,\perp}(808 \ nm)} \approx 100.
\end{equation}

This high polarization contrast gives us therefore an insight into the absorption efficiency achievable in this nano-absorber upon control of the incident laser polarization. The absorption efficiency for the parallel case is $Q_{abs,\parallel} \approx 3.64$, while for the perpendicular case $Q_{abs, \perp} \approx 0.03$ (the area of the nanorod being $A_{nr} \approx 3.29 \cdot 10^{-16} \ m^2$, whose calculation is based on the sizes extracted from FEM simulations). The measurements have been compared also to the FEM simulations (black dots), showing a good match. It is worth noting how the measurements follow the typical pattern $\sigma_{abs}(\lambda, \theta_{pol}) \ = \ \sigma_{abs, \parallel}(\lambda)cos^2(\theta_{pol})$ expected for perfect dipoles \cite{Ming2009, Muskens2008, Heylman2016}. 

For comparison, Fig.~\ref{fig:polar plot snr} shows also the nanomechanical photothermal signal at $\lambda_{exc}$ = 513 nm (the blue dots). The TSPR shows almost no polarization contrast since it starts to overlap with polarization-independent electronic transitions in gold \cite{Johnson1972}. For this reason, the plasmonic damping increases, resulting in an transverse plasmonic resonant absorption strength smaller than the longitudinal one.

Nanomechanical photothermal spectromicroscopy also allows the precise determination of the orientation on the substrate of different nano-absorbers, as seen in Fig.~\ref{fig:2D scans}. 2D maps of the same region on the drum resonator for three different polarization angles $\theta_{pol}$ (90°, 135°, 180°) are acquired at $\lambda_{exc} = 808$ nm. The two responses are from two individual nanorods, whose absorption amplitude varies as a function of the laser polarization. Focusing the attention on the two perpendicular polarizations, 90° and 180°, the two absorbers behave in an opposite way, meaning that they are almost perpendicular to one another, while absorbing almost the same amount of light for the central scenario ($\theta_{pol} = 135$°). Nanomechanical photothermal spectromicroscopy can be therefore employed in the analysis of the more complex optical features, for nanorods and more exotic structures, like plamon-assisted optical chirality in metallic nanoparticles \cite{Spaeth2019, Spaeth2022, Rangacharya2020}.

\begin{figure}
    \begin{center}
    \centering
    \includegraphics[width = 0.5\textwidth]{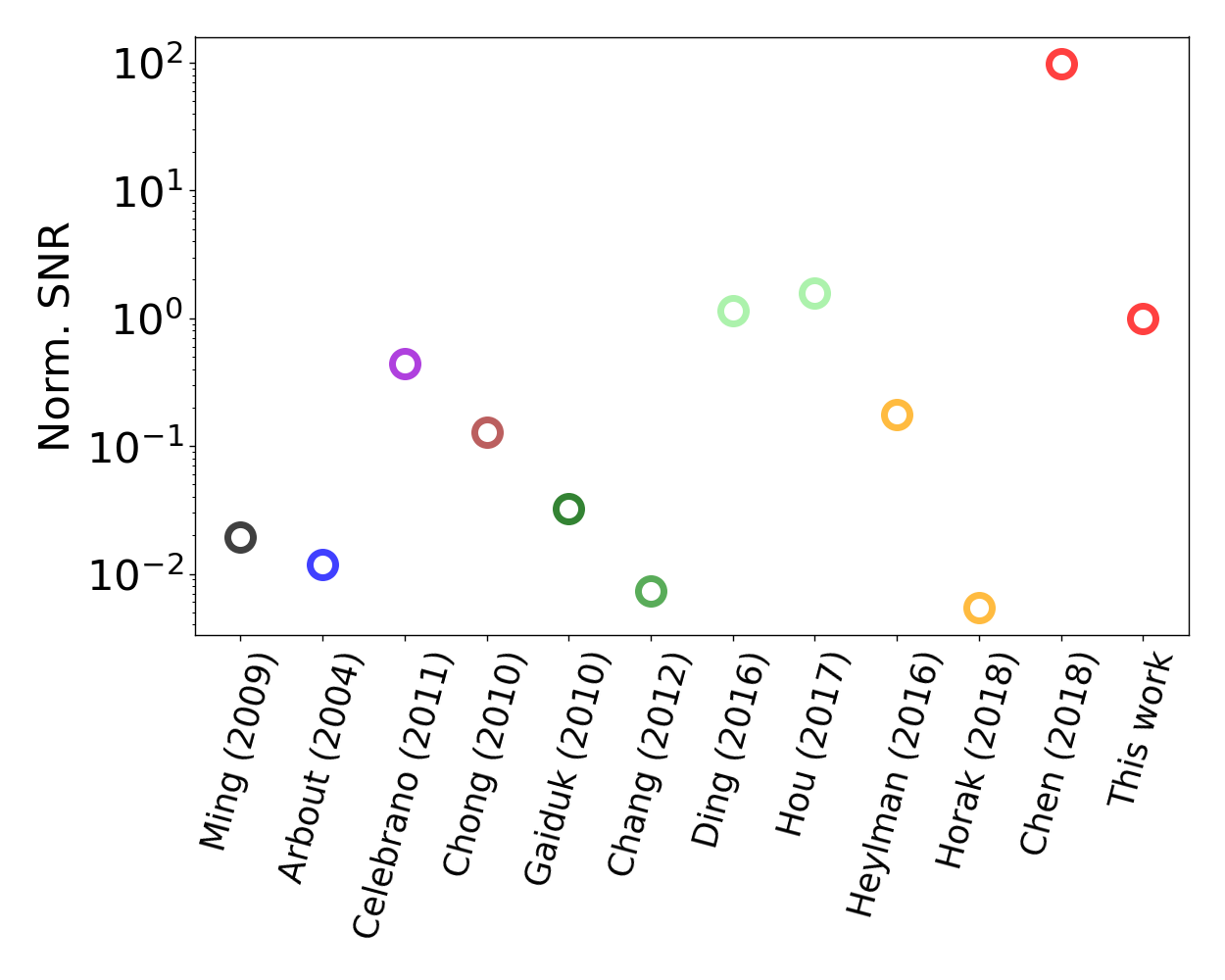}
    \caption{Signal-to-noise ratio (SNR) comparison between different single-molecule absorption sensing techniques. Black: UV-Vis exctinction \cite{Ming2009}; blue: spatial modulation spectroscopy \cite{Arbouet2004}; violet: extinction microscopy plus balance detection \cite{Celebrano2011}; brown: ground-state depletion microscopy \cite{Chong2010}; dark green: photothermal constrast microscopy (PCM) with Glycerol \cite{Gaiduk2010}; green: PCM with thermotropic liquid crystal 5CB \cite{Chang2012}; light green: PCM with near-critical Xe \cite{Ding2016, Hou2017}; orange: optical microresonator \cite{Heylman2016, Horak2018}; red: nanomechanical photothermal microscopy \cite{Chien2018}.}
    \label{fig:SNR}
    \end{center}
\end{figure}

Finally, to further stress the advantages offered by nanomechanical photothermal spectromicroscopy compared to other label-free single-particle and molecule spectromicroscopy techniques, a comparison between SNR of different approaches is carried out in the following way\cite{Goldsmith2021}:
\begin{equation} \label{eq5}
    Norm. SNR = \frac{SNR_{exp}}{SNR_{0}} \frac{P_{heat,0}}{P_{heat, exp}} \sqrt{\frac{\tau_{m,0}}{\tau_{m,exp}}}.
\end{equation}
For the sake of completeness, Eq.~\ref{eq5} takes into account, together with the SNR itself, also the power absorbed by the sample under study $P_{heat}$ and the time constant of the experiment $\tau_m$. The reference values of the three quantities used for normalization (labelled with the subscript $0$) corresponds to the value obtained for the individual nanorod of Fig.~\ref{fig:single nr}\&\ref{fig:polar plot snr}. The calculations are plotted in Fig.~\ref{fig:SNR}, for which the experimental values extracted from the listed references are used (for the used values, see SI) \cite{Ming2009, Arbouet2004, Celebrano2011, Chang2010, Chang2012, Ding2016, Hou2017, Heylman2016, Horak2018, Chien2018}.

The different colors correspond to different techniques (see caption). It is worth noting that the nanomechanical photothermal approach of Ref.~\citenum{Chien2018} (first red empty dot) presents the highest SNR, followed by photothermal contrast microscopy with near-critical Xe \cite{Ding2016, Hou2017} and this work. More precisely, the difference of two orders of magnitude between this work and Ref.~\citenum{Chien2018} lies in the different prestress of the nano-optomechanical resonator. There, oxygen plasma treatment is exploited to reduce the stress of the resonator, with the aim to improve its relative power responsivity $R_P$ to detect single Atto 633 molecules. Here, there has been no need to further reduce the stress due to the already high sensitivity of the resonator used for the nanorods detection. Still, this work shows the superior capabilities of nanomechanical photothermal spectromicroscopy over a huge range of label-free absorption techniques. 

\section{Conclusions}

In conclusion, the optical absorption cross-section of individual silica-coated gold nanorods in the NIR range has been measured and quantitatively characterized using nanomechanical photothermal spectroscopy and microscopy, likewise elucidating their polarization features. With this approach, where the substrate acts as a temperature sensor, it is possible to shed light on the variations in nano-absorbers' properties to investigate concealed heterogeneity, as expected for these complex systems, as well as their reciprocal intercoupling, which opens up a wealth field of research by its own. It has also been shown that these nanorods present, on one hand, a pronounced plasmonic electron-surface scattering, broadening their LSPR in conjunction with bulk scattering. On the other hand, a strong polarization contrast of the order of few hundreds has been observed. The interaction between the silicon nitride slab and the nanorod has been also investigated, consisting in a modulation of its absorption strength over the whole considered spectrum, while weakly affecting the plasmonic resonant energy and its broadening. This result underlines the importance of taking into account the interaction of the substrate in all the experiments where a support is used for spectroscopic measurements. 

Primarily, this work demonstrates the capabilities of nanomechanical photothermal NIR spectromicroscopy for localizing individual nanoparticles, obtaining their plasmon spectra, and resolving their polarization features, pushing further our understanding of light-matter interaction at the nanoscale level.
A comparison conducted among the different label-free single-molecule techniques shows that nanomechanical photothermal sensing presents a superior signal-to-noise ratio within a less complex experimental setup and measurement procedure.

\begin{suppinfo}

The following files are available free of charge.
\begin{itemize}
  \item Supporting Information: FEM simulations details (computational approach, gold dielectric function model), LSPR linewidth (measurements and modeling), FEM substrate analysis, table with all the parameters used for the SNR comparison (PDF)
\end{itemize}

\end{suppinfo}



\begin{acknowledgement}

The authors are thankful to Sophia Ewert and Patrick Meyer for the device fabrication, and for the assistance of Sophia Ewert with scanning electron microscopy. The authors thank also Hendrik Kähler and Andreas Kainz for the fruitful discussions on finite element method. Furthermore, the authors thank Johannes Hiesberger, Paolo Martini and Niklas Luhmann for useful discussions on the set-up. Finally, K.K. thanks BioRender for their illustration software. This work is supported by the European Research Council under
the European Unions Horizon 2020 research and innovation
program (Grant Agreement No. 716087-PLASMECS). 

\end{acknowledgement}

\bibliography{bibliography_main.bib}


\end{document}


\section{FEM nanorod simulations}

\begin{figure}
    \begin{center}
    \centering
    \includegraphics[width = 1\textwidth]{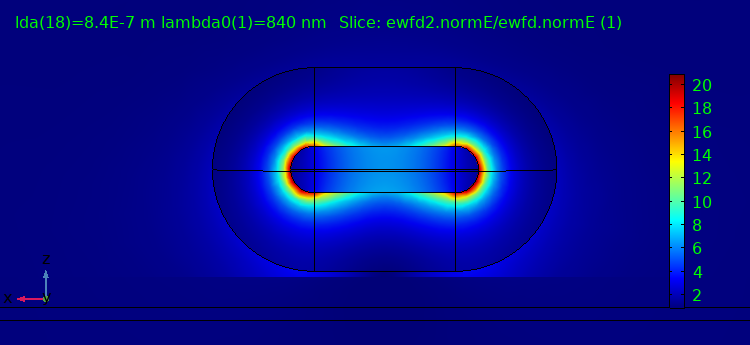}
    \caption{Simulated field enhancement assuming a laser beam with $\lambda = 840$ nm polarized parallel to the long axis of the nanorod. It takes place in the region surroundings the hemispherical caps of the nanorod, of the order the particle diameter \cite{Davletshin2012, Tian2012}.}
    \label{fig:FEM FE}
    \end{center}
\end{figure}

The finite element model of the silica-coated gold nanorods lying on a silicon nitride substrate was built using the commercially available FEM software COMSOL Multiphysics (COMSOL Inc., Burlington, MA), version 5.5. The shape of the nanorods was extracted from SEM measurements, and each single nanorod was mimicked by a cylinder of length $L_{nr}$, radius $r_{nr}$, terminated with hemispherical caps with the same radius (see Fig.\ref{fig:FEM FE}) \cite{Lee2005}. In this way, concerning the nanorod gold core, the only parameters to be changed for comparison between measurements and simulations are the length $L_{nr}$ and the radius $r_{nr}$. It is worth mentioning that in prolate ellipsoids, the aspect ratio plays a central role in determining the wavelength $\lambda_{SPR}$ and amplitude $\sigma_{abs}(\lambda_{SPR})$ of the longitudinal SPR~\cite{Muskens2008}.

\begin{figure}
    \begin{center}
    \centering
    \includegraphics[width = 0.75\textwidth]{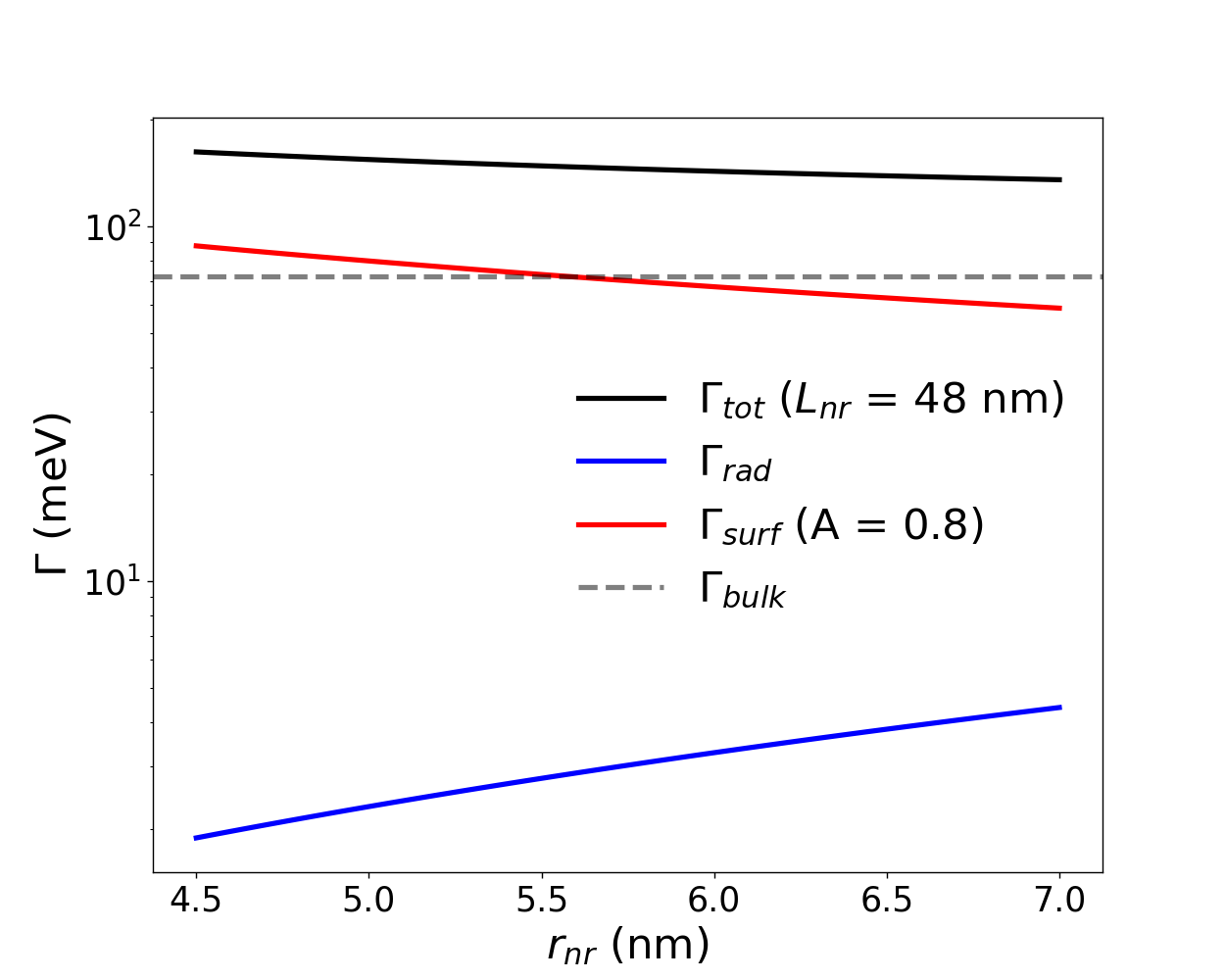}
    \caption{Analytical model of the damping mechanisms behind the broadening of the nanorod LSPR resonance. $\Gamma_{bulk}$ represents the mean bulk-like electron scattering rate (grey dashed line); $\Gamma_{surf}$ represents the electron-surface scattering rate (red solid line); $\Gamma_{rad}$ represents the radiative damping (blue solid line). The black solid line indicates the overall damping $\Gamma_{tot}$, sum of the three contributions. Electron-bulk and surface scattering are the major source of plasmonic damping, with the radiative one being almost two orders of magnitude smaller.}
    \label{fig:damping}
    \end{center}
\end{figure}

The FEM simulation resolves the spatial distribution of the electromagnetic field on the defined physical domain, \textbf{E}(\textbf{r}), using both the full and scattered field formulation in two successive steps \cite{Knight2009, Comsol2022}. In the first step, the interaction between an incident electromagnetic plane wave and the substrate is computed with the full field formulation in the absence of the nanorod. The incident electric field is set to be parallel to the nanorod's long axis. The results of this simulation are then used as input background electric field for the second step, where the scattered field formulation is exploited to calculate the overall electric field inside the nanorod and its corresponding absorption cross-section. Perfectly matched layers (PML) are used to avoid artificial reflections at the simulation domain boundaries and act as an anisotropic absorbing layer. Both PML and mesh setting parameters are optimized to show very good agreement between spherical gold nanoparticles and Mie theory predictions in free space, both for the absorption and scattering cross-sections. In Fig.~\ref{fig:FEM FE} the electric field enhancement in the nanorod and its surroundings is plotted, showing how the electric field is strongly concentrated at the nanorod's tips for the first order LSPR, giving rise to the so-called hotspots. Once the electromagnetic field is computed, the nanorod's absorption cross-section is obtained by integrating the absorbed power density due to resistive losses $Q_h$ over the nanorod's volume $V_{NP}$ (including both the gold core itself and its silica coating), and dividing it by the incident planewave light intensity $I_0$ \cite{Grand2020}:
\begin{equation} \label{eq5}
    \sigma_{abs}(\omega) = \frac{1}{I_0} \iiint_{V_{NP}}{\frac{1}{2}\omega\epsilon_0Im(\epsilon_{NR}(\omega, \textbf{r}))|\textbf{E}(\textbf{r})|^2}\;d\textbf{r}
\end{equation}
$\epsilon_0$ is the vacuum dielectric constant and $Im(\cdot)$ the imaginary part operator applied here on the dielectric function. The dielectric response of the gold core is defined starting from the bulk values for gold, $\epsilon_{bulk}$ (taken from~\citenum{Johnson1972}), and corrected taking into account the intrinsic size effects. Here, both electron-surface scattering and radiative damping were introduced on the overall complex dielectric function for the metal core $\epsilon_{NR}$:
\begin{equation} \label{eq6}
    \epsilon_{NR}(\omega, L_{eff}) = \epsilon_{bulk} + \frac{\omega_p^2}{\omega^2 + i\omega\gamma_0} - \frac{\omega_p^2}{\omega^2 + i\omega(\gamma_0 + \frac{A v_F}{L_{eff}} + \frac{\eta V}{\pi})}.
\end{equation}
Here, $\gamma_0$ represents the bulk-like electron scattering rate for gold; $v_F$ indicates the electron Fermi velocity; A is a dimensionless parameter related to the details of the electron surface scattering; $L_{eff}$ the electron mean free path confined at the surface, dependent on the sizes of the nanorod \cite{Davletshin2012, Nima2017}. Finally, V is the nanorod volume and $\eta$ represents an effective radiation damping rate \cite{Nima2017}. Therefore, the resulting broadening of the LSPR can be modelled as:
\begin{equation} \label{eq7}
    \Gamma_{tot} = \Gamma_{bulk} + \Gamma_{surf} + \Gamma_{rad}
\end{equation}
with $\Gamma_{bulk}$ = $\gamma_0$, $\Gamma_{surf}$ = $A v_F$/$L_{eff}$ and $\Gamma_{rad}$ = $\eta V$/$\pi$. The parameter A is another degree of freedom used for the comparison between measurements and simulations. For the specific sizes investigated here, the radiative damping is almost two orders of magnitude smaller than the surface and bulk contributions (see Fig.~\ref{fig:damping}). This is due to the reduced volume compared to gold nanoparticles of the same sizes, for which it starts to be the main damping mechanism for radii > 30 nm. It is also worth noting, on one hand, that for a fixed value of A and nanorod's length, the overall damping shows a weak dependence on the nanorod's radius in the available range of values. On the other hand, for a fixed set of sizes, the plasmonic linewidth shows a strong dependence on the surface scattering parameter, A (see next section).

\section{LSPR linewidth and difference between individual nanorods and aggregation}

\begin{figure}
  \begin{center}
  \centering
  \subfloat[\centering]{\includegraphics[width = 0.50\textwidth]{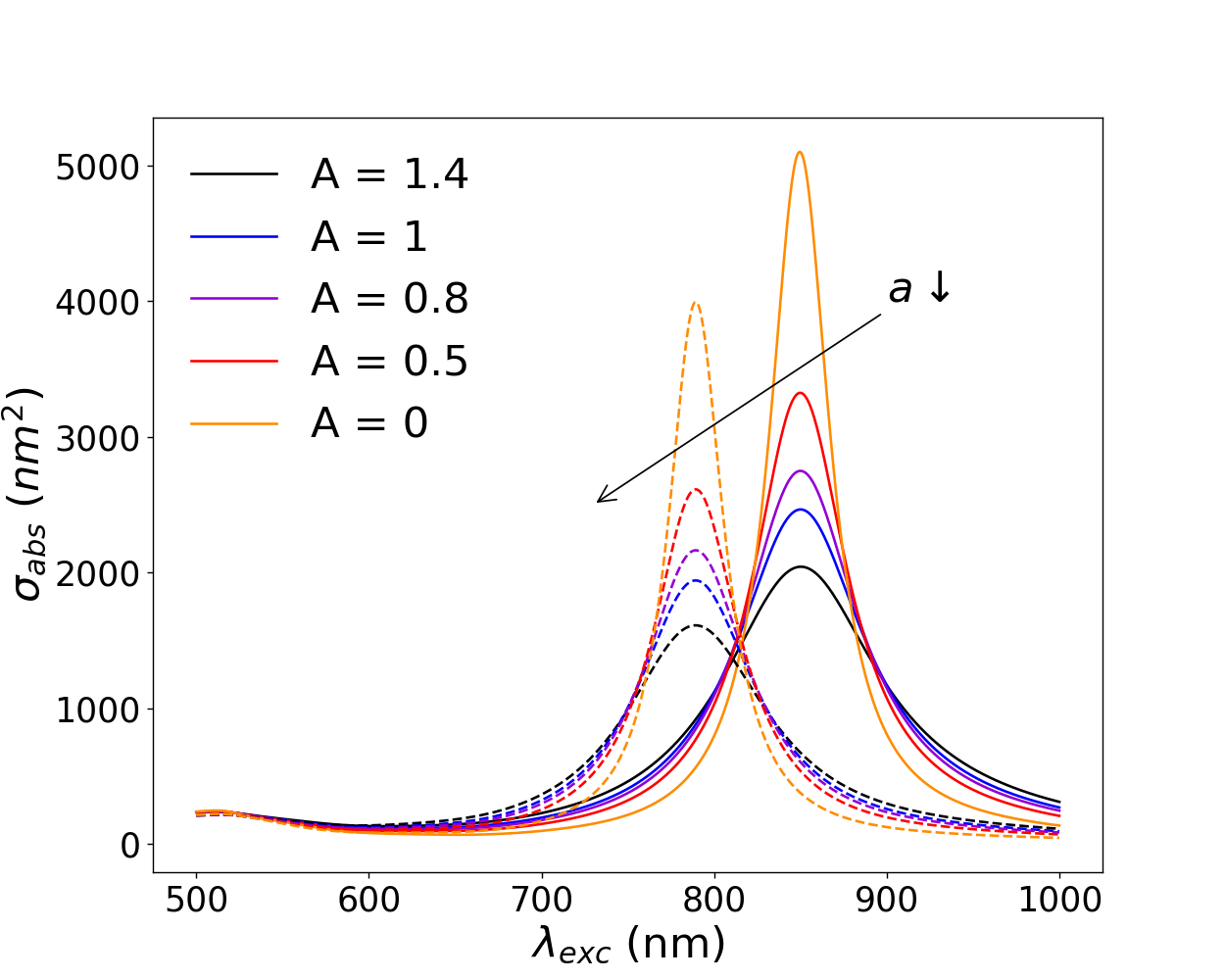} \label{fig:snr spectra tmatrix}}
  \subfloat[\centering]{\includegraphics[width = 0.50\textwidth]{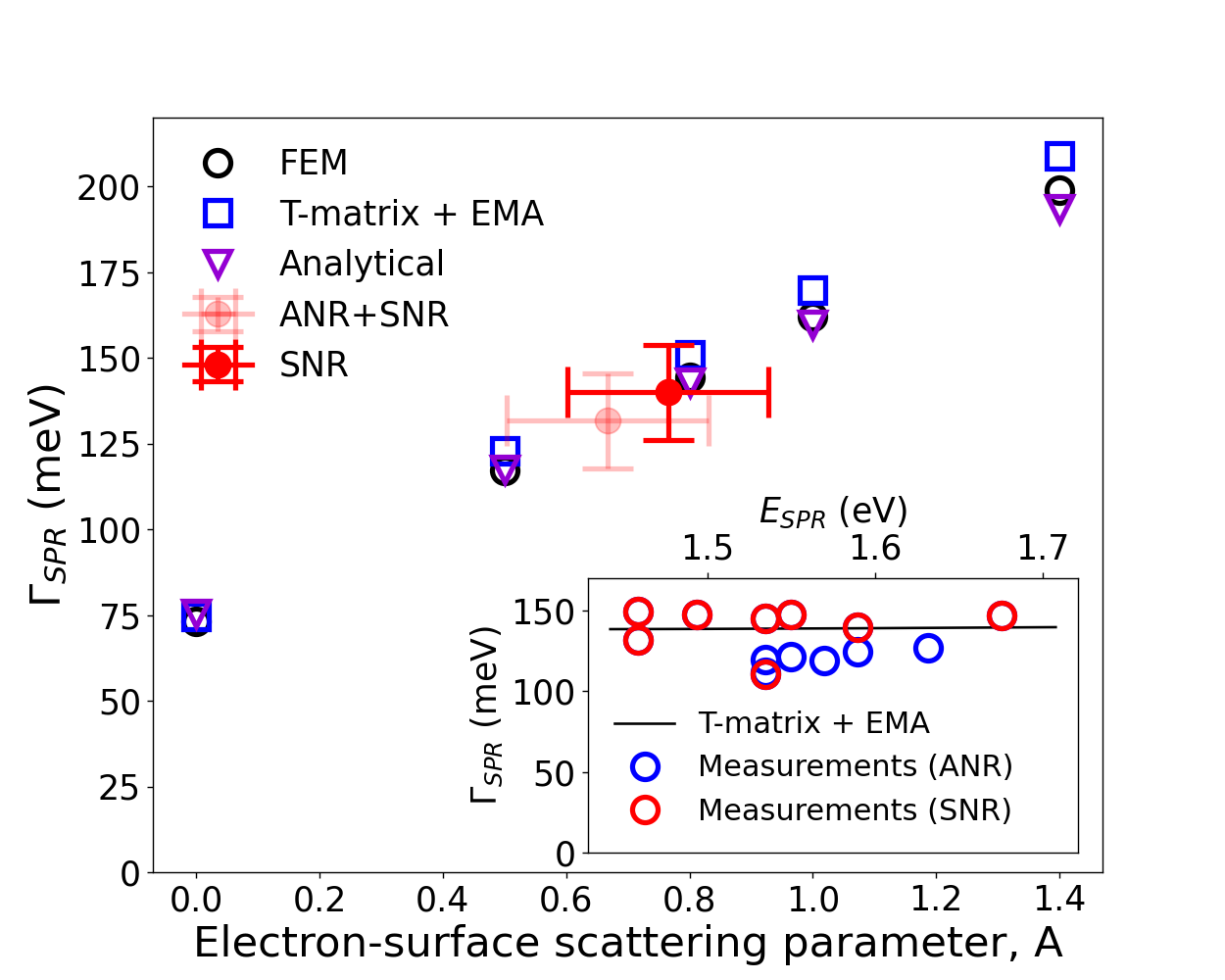} \label{fig:linewidth}}
  \caption{(a) Absorption spectra of the LSPR resonance calculated with the T-matrix method together with effective medium approximation (EMA), for different value of the parameter A, for two different aspect ratios $a$ (3.5, dashed lines; 4, solid lines). (b) LSPR linewidth's dependence on the parameter A for: the analytical model (eq. \eqref{eq7}), empty dark violet downward triangles; FEM, empty black circles; T-matrix plus EMA, empty blue squares; mean value and standard deviation of the measured linewidths considering single nanorods, red dot, and single nanorods plus aggregations, light red dot. Inset: linewidth as a function of the LSPR energy for the T-matrix calculation (black solid line) and measurements of single nanorods (empty red dots), and aggregations (empty blue dots).} \label{fig:exp snr}
  \end{center}
\end{figure}

The nanorods LSPR resonances are measured and their linewidths quantified. These values are then compared with the analytical model introduced above in Eq.~\eqref{eq7}, together with the FEM simulations, and the T-matrix method \cite{pySCATMECH2022} applied with the effective medium approximation (EMA) \cite{Lioi2022}, for the same nanorod's shape as used in the FEM analysis. The last approach used here is not thought to be fully explanatory for the nanorod's LSPR resonances analyzed in the main text (for which FEM simulations were carried out), but it helps to understand in a simple way how the absorption spectrum changes as function of the electron-surface scattering and to differentiate between single gold nanorods and aggregations of two or more nanoabsorbers. As for the FEM, the T-matrix method requires as input the material parameters of gold (modelled exactly as for the FEM with Eq.~\eqref{eq6}). The silica coating is taken into account by considering an effective dielectric function of the medium surrounding the nanorod, as done by Lioi et al. \cite{Lioi2022}. Fig.~\ref{fig:snr spectra tmatrix} shows how strongly the LSPR resonance depends on the surface scattering parameter A, for two different aspect ratios $a \ = \ L_{nr}/(2\ r_{nr})$ (3.5 for the dashed lines, 4 for the solid lines). It is worth noting that already going from a value of 0 (bulk plus radiative damping) to a value of 0.5, a strong reduction of the absorption peak $\sigma_{abs}(\lambda_{SPR})$ is observable in both cases (35$\%$ reduction). The higher the value of A, the lower the absorption peak. It is also worth noticing that by reduction of the nanorod's aspect ratio $a$, the LSPR energy is blue-shifted \cite{Brioude2005}, increasing the overlap between the plasmonic-assisted resonance and the interband transitions, ultimately resulting in a reduction of the overall absorption cross-section for the same values of electron-surface scattering. The plasmonic linewidths measured in this study lie in a range comprised between ca. 110 and 150 meV, meaning that electron-surface scattering must be accounted for. By application of the analytical model on the measured plasmonic resonances, a value of A is extracted for each measurement, covering a range between 0.4 and 0.9, as seen in Fig.~\ref{fig:linewidth}. The spread could be due to non-uniform distribution of the silica coating, as well as surface defects of the gold core itself. Interestingly the linewidth follows linearly the change of the parameter A (blue empty triangles), allowing a simple a evaluation for the experimental values. The same behavior is observed with the FEM simulations (empty black dots) and with the T-matrix method (empty blue squares). It is also shown once again the better matching of the FEM with the analytical model compared to the T-matrix, demonstrating once again the superiority of the FEM approach. The light red dot indicates the average value of the measured linewidths (Fig.~\ref{fig:linewidth}, inset), considering both single and aggregated nanorods, together with the standard deviations computed both for the linewidth itself and the parameter A, showing the spread of the obtained values. With an average value of 0.65 for A, one can conclude that all the spectrally resolved nanorods are strongly affected both by electron-bulk and surface scattering. With this in mind, it is therefore clear that also the spread of values for the absorption cross-section peak of a single nanorod will be limited. Focusing the attention again on Fig.~\ref{fig:snr spectra tmatrix}, it is visible that the absorption peak cannot go beyond $4 \cdot 10^{-15} \ m^2$ for $a \ = \ 3.5$, and beyond $5 \cdot 10^{-15} \ m^2$ for $a \ = \ 4$, even neglecting the plasmonic surface scattering. As already stated above, this threshold increases with increasing aspect ratio. It could be therefore assumed that, given the measured plasmonic linewidths, all the absorption peaks roughly > $3 \cdot 10^{-15} \ m^2$ for $\lambda_{SPR}$ < 800 nm, and > $3.5\cdot10^{-15} \ m^2$ for $\lambda_{SPR}$ > 800 nm, are due to aggregations of two or more nanorods, rather than a single one.

\section{Silicon nitride substrate}

\begin{figure}
    \begin{center}
    \centering
    \includegraphics[width = 1\textwidth]{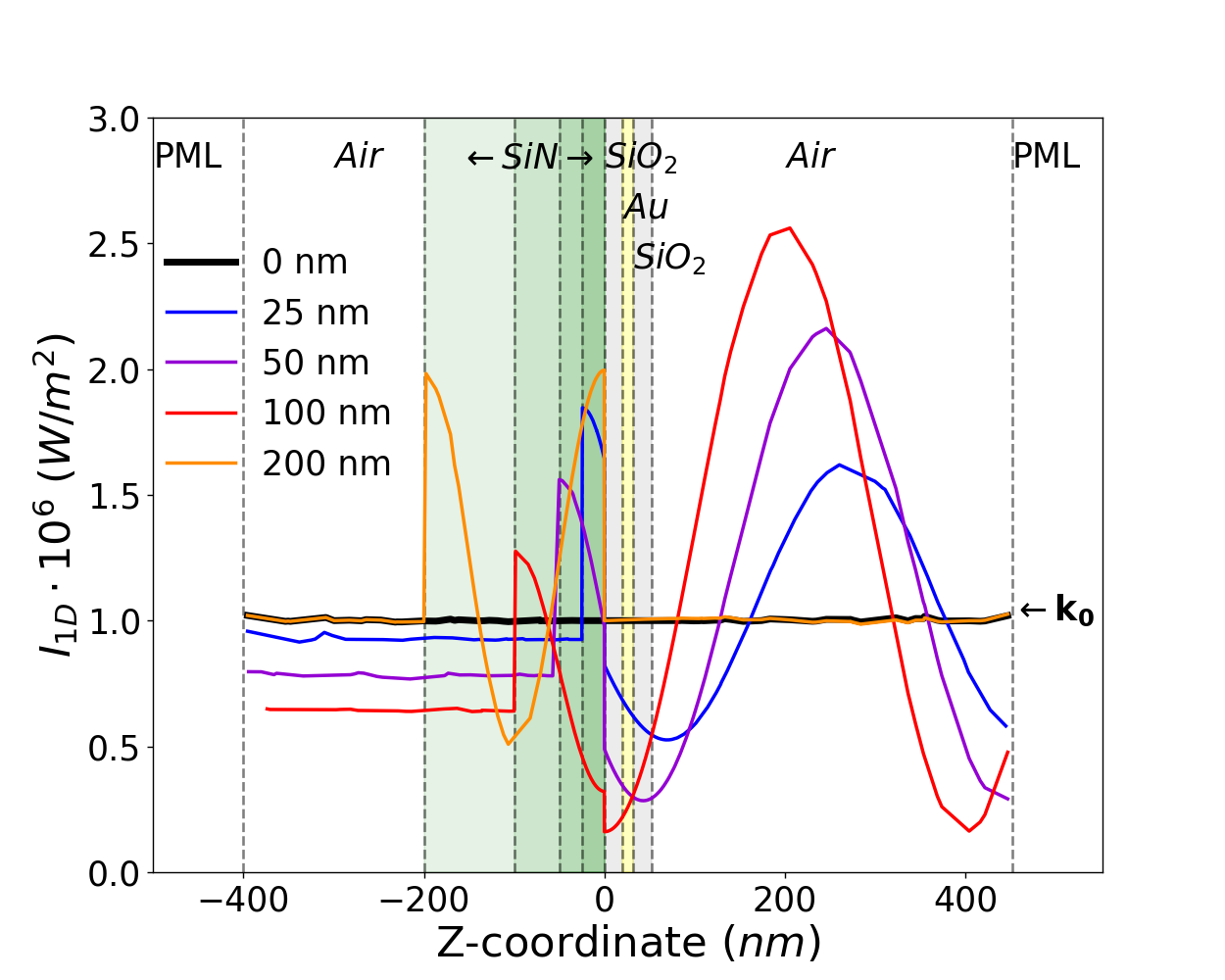}
    \caption{FEM simulated intensity distribution along a 1D cut line passing in the center of the physical domain at a wavelength of 800 nm, for different silicon nitride slab thicknesses. The intensities are the results to the FEM first step, where only the slab is simulated, without any gold nanorod on top of it. The vertical lines and the relative colors show the positions of each element along the cut line: white, air; grey, silica; yellow, gold; green of different intensities, the different silicon nitride slabs. }
    \label{fig:I substrate}
    \end{center}
\end{figure}

FEM simulations have been performed to study also how to the substrate affects the absorption cross-section spectrum for a individual gold nanorod. This analysis is carried out by evaluating the absorption cross-section for different thicknesses of the silicon nitride slab placed underneath the nano-absorber (see Fig.~3b in the main text). For sake of simplicity, the study is performed for one single wavelength (here 800 nm), but it can be extended to the whole spectrum considered in this work. Fig.~\ref{fig:I substrate} shows the intensity distribution at 800 nm wavelength along a cut line parallel to the optical axis, passing in the center of the simulated physical domain. The intensities refer to the FEM first step, where only the silicon nitride slab is taken into account and the individual nanorod is not considered. Here, the full field formulation is used, i.e. an input electric field is defined on an input port on the simulated domain, and the total electric field, sum of the input and the refracted field, is calculated. In this way, it is possible to understand how the light interference due to the presence of the substrate affect ultimately the absorption cross-section of the nano-absorber. The vertical lines and the related colors, together with the tags, are plotted to indicated the positions occupied by each element along the central cut line in the physical domain. Even if not present in the simulation first step, the gold core and the silica coating of the nanorod are included for sake of clarity. The input electric field (indicated by $\bf{k_0}$ in the plot) enters the domain from the right-side (Z-coordinate positive values) and propagate towards the Z-coordinate negative values. Of particular interest is how the full field intensity shows different interference patterns for different silicon nitride slab thicknesses. Due to this effect, the intensity distribution in the proximity of the nanorod assumes different values for the different cases. For instance, going from 0 nm thickness (black curve) to 100 nm (red curve) the intensity in the nanorod's position decreases (yellowish region of z-values), resulting in a reduction of the absorption cross-section at 800 nm wavelength. For a thickness of 200 nm (dark orange curve), the silicon nitride slab acts as if it were transparent to this wavelength, resulting in a intensity distribution outside the slab almost identical to the case where no substrate is present at all. This is due to the value assumed by the refractive index $n$ of the silicon nitride at this wavelength (the spectral distribution of the silicon nitride index of refraction has been taken from \citenum{Philipp_1973}). For a vacuum wavelength of $\lambda_0$ = 800 nm, $n \ \approx \ 2$ inside the slab, which results in an effective wavelength $\lambda_n$ = $\lambda_0$/$n$ $\approx$ 400 nm in the silicon nitride region, equivalent to half of the substrate thickness for the last curve in Fig.~\ref{fig:I substrate} (dark orange). Therefore, half of the wavelength is contained in the slab, making it transparent to the incident electromagnetic wave.  In this way, the absorption cross-section is going to be almost of the same magnitude as for the free space case. Indeed, the same electric field will enter the Eq.~\eqref{eq5}, resulting in equal electromagnetic energy losses. Consequently, a periodic pattern is recovered between the absorption cross-section at a specific wavelength and the thickness of the substrate, as it has been shown in Fig.~3b in the main text.

\section{Signal-to-Noise ratio comparison}

With the advantage of a reduction in experimental complexity compared to other label-free techniques, nanomechanical photothermal spectromicroscopy offers a highly sensitive approach to measure single particle and molecule. To show that, a comparison of the signal-to-noise ratio (SNR) among different label-free single-molecule technique is carried out (see the main text) \cite{Goldsmith2021}. The three main quantity used to have a meaningful comparison are: the experimental SNR itself $SNR_{exp}$, as given by the authors (when not explicitly stated, averaged values are used instead); the power absorbed by the sample under study $P_{heat}$; the time required to perform a meaningful experiment $\tau_m$. Table~\ref{table:1} summarizes all the useful information for the calculations developed in the main text. Each technique has been referenced with the corresponding work from which the experimental values have been extracted. Together with the three aforementioned quantities, Table~\ref{table:1} displays also the type of sample analyzed, as well as its corresponding absorption cross-section $\sigma_{abs}$ and the pump intensity used to stimulated it (taking $1/e^2$ as definition of the laser beam waist). For sake of clarity, some of the techniques taken into account in Table~\ref{table:1} could performed only microscopy studies at the time of publication of the referenced works. Exception done for ground-state depletion microscopy, all the other techniques started to offer also spectroscopy capabilities in the last years.

\begin{table}
\centering
\caption{Parameters of the different techniques for the SNR comparison. SMS: spatial modulation spectroscopy; GSD: ground-state depletion microscopy; PCM: photothermal constrast microscopy; OMM: optical microresonator microscopy; NPM: nanomechanical photothermal microscopy; NPSM: nanomechanical photothermal spectromicroscopy. NG: not given. *: at the time of publication. **: assumed values.}
\begin{tabular}{ |p{2cm}|p{2.2cm}|p{1.84cm}|p{2cm}|p{1.7cm}|p{1.25cm}|p{1.25cm}|p{1.3cm}|  }
 \hline
 \textbf{Technique} & \textbf{Capability*} & \textbf{Pump Intensity ($kW/cm^2$)} & \textbf{Sample} & $\sigma_{abs}$ ($m^2$) & $P_{heat}$ ($pW$) & $\tau_m$ ($ms$) (avgs)  & $SNR_{exp}$ \\
 \hline
 UV-Vis\, Extinc. \cite{Ming2009} & Spectro-\,microscopy & NG & Nanorod & NG & 1274** & 208** & 6.38**\\
 \hline
 SMS \cite{Arbouet2004} & Spectro-\,microscopy & 22 & Metal cluster & 4.3 $\cdot 10^{-16}$ & 93620 & 10000* & 2000\\
 \hline
 Extinc. + Bal.Det. \cite{Celebrano2011} & Microscopy & 280 & TDI dye &  1.8 $\cdot 10^{-19}$ & 508 & 2 (10x) & 5.7\\
 \hline
 GSD \cite{Chong2010} & Microscopy & 590 & Atto dye & 5 $\cdot 10^{-20}$ & 294 & 30 (20x) & 3.7\\
 \hline
 PCM \,(Glycerol) \cite{Gaiduk2010} & Microscopy & 9300 & BHQ & 4 $\cdot 10^{-20}$ & 1000 & 300 & 10\\
 \hline
 PCM (5CB) \cite{Chang2012} & Microscopy & 28  & Nanosphere & 4.8 $\cdot 10^{-16}$ & 132000 & 20 & 78\\
 \hline
 PCM (Xe) \cite{Ding2016} & Microscopy & 28 & Nanosphere & 4.8 $\cdot 10^{-16}$ & 64 & 50 & 9.4\\
 \hline
 PCM (Xe) \cite{Hou2017} & Microscopy & 0.45 & CP & 4 $\cdot 10^{-18}$ & 64 & 30 & 10\\
 \hline
 OMM \cite{Heylman2016} & Spectro-\,microscopy & 2 $\cdot 10^{-4}$  & Nanorod & 1 $\cdot 10^{-14}$ & 20 & 1000 (30x) & 2\\
 \hline
 OMM \cite{Horak2018} & Spectro-\,microscopy & 522 & CP & 8 $\cdot 10^{-19}$ & 4100 & 100 & 4\\
 \hline
 NPM \cite{Chien2018} & Microscopy & 35.4 & Atto dye & 4.8 $\cdot 10^{-20}$ & 6.3 & 40 & 70\\
 \hline
 NPSM & Spectro-\,microscopy & 4.98 & Nanorod & 2.5 $\cdot 10^{-15}$ & 120947 & 200 & 30759\\
 \hline
\end{tabular}
\label{table:1}
\end{table}

\newpage

\bibliography{bibliography_main.bib}